\documentclass[fleqn,twoside]{article}
\usepackage{espcrc2}

\usepackage{graphicx}
\usepackage[figuresright]{rotating}

\newcommand{\AmS}{{\protect\the\textfont2
  A\kern-.1667em\lower.5ex\hbox{M}\kern-.125emS}}

\title{The quark mass and $\mu$ dependence of the QCD chiral critical point
        \thanks{Presented by Ch. Schmidt. The work has been supported by the
        DFG under grant FOR 339/1-2 and by PPARC grant PPA/A/S/1999/00026.}}

\author{Ch. Schmidt\address[Bi]{Fakult\"at f\"ur Physik, Universit\"at Bielefeld,
        D-33615 Bielefeld, Germany},
        C.R. Allton\address[Swan]{Department of Physics, University of Wales
        Swansea, Singleton Park, Swansea, SA2 8PP, U.K.},
        S. Ejiri\addressmark[Swan],
        S.J. Hands\addressmark[Swan],
        O. Kaczmarek\addressmark[Bi],
        F. Karsch\addressmark[Bi] and
        E. Laermann\addressmark[Bi]}
\begin{document}

\begin{abstract}
In order to study the QCD chiral critical point we investigate Binder
Cumulants of the chiral condensate. The results were obtained from simulations
of 3 and 2+1 flavors of standard staggered fermions and 3 flavors of p4
improved staggered fermions. The quark masses used are close to the
physical quark mass. To extract the dependence on quark mass and chemical
potential we apply a new reweighting technique based on a Taylor expansion of
the action. The reweighting accuracy is ${\cal O}(m)$ for the standard and
${\cal O}(m^2)$, ${\cal O}(\mu^2)$ for the p4 action.
\vspace{1pc}
\end{abstract}

\maketitle

\section{INTRODUCTION}
\begin{figure}[htb]
\includegraphics[width=7.5cm, height=5cm]{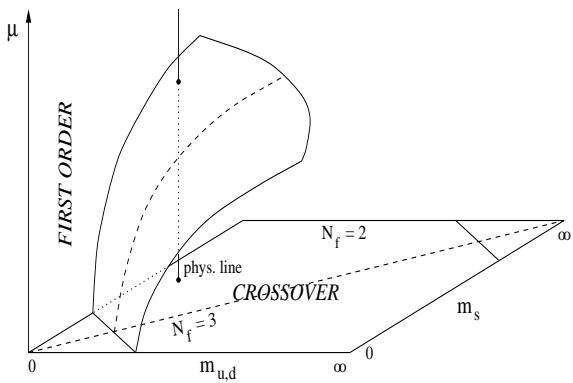}
\caption{Generic phase diagram of QCD
%Diagram of the order of the QCD phase transition, in
%  space of chemical potential and quark masses.
}
\label{fig:phase_diagram}
\end{figure}

The phase transition in 3-flavor QCD is first order in the chiral limit
($m\equiv0$). This first order transition will persist for small but non-zero
values of $m$ up to a critical value, $\bar{m}$, of the quark mass. QCD at this
chiral critical point belongs to the universality class of the 3d Ising model
\cite{Gav94,schmidt}. The generic phase diagram of QCD is shown in
figure~\ref{fig:phase_diagram}.
In the space of two light quark masses $m_{u,d}$, one heavier quark mass $m_s$
and the quark chemical potential $\mu$, one expects a critical surface, which
bends over the quark mass plane and separates the regime of first order phase
transitions from the crossover regime.

To probe the order of the phase transition we compute the Binder Cumulant of
the chiral condensate $B_4$ at $\mu=0$ for several quark masses at the
corresponding critical coupling $\beta_c(m)$. Monte Carlo simulations for
$\mu\ne 0$ are not possible due to the sign problem. $B_4$ is given by
\begin{equation}
B_4={\left<(\delta\bar\psi \psi)^4\right>_{\beta_c,m} \over
  \left<(\delta\bar\psi \psi)^2\right>_{\beta_c,m}^2}\, ,
\quad
\delta\bar\psi \psi = \bar\psi \psi - \left<\bar\psi \psi\right>\, .
\end{equation}
This quantity is a renormalization group invariant quantity for $m=\bar{m}$
with a universal value, which depends on the universality class only. For the
3d Ising model we have $B_4=1.604$ \cite{b4ising}. The critical surface is
given by the surface of constant $B_4=1.604$. We define the critical coupling
$\beta_c$ as the peak position of the chiral susceptibility.
%$\left<(\delta\bar\psi \psi)^2\right>$.

\vspace*{-19.2cm} \mbox{} \hfill {\large BI-TP 2002/22} \\
                \mbox{} \hfill {\large  SWAT/347}     \\
\vspace*{18.2cm}
\section{REWEIGHTING IN $m$ AND $\mu$}
In order to compute lines within the critical surface, we apply a reweighting
technique based on a Taylor expansion \cite{allton,ejiri}. Reweighting in
couplings $w=m,\mu$ which couple to fermionic operators, is possible by
introducing a reweighting factor $R$ in the expectation value
\begin{equation}
\left<O\right>_{w}=\left<OR(w,w_0)\right>_{w_0} / \left<R(w,w_0)\right>_{w_0}.
\end{equation}
Here the reweighting factor is given by
\begin{equation}
\ln R= \ln \det Q(w) - \ln \det Q(w_0),
\end{equation}
with the fermion matrix $Q$. After expanding $\ln R$ one is left with
\begin{equation}
\ln R = \sum_{n=1}^{\infty}{(w-w_0)^n\over n !} R_n.
\end{equation}
For fermionic observables like $\bar\psi \psi$ it is also necessary to expand
the observable itself. For details see \cite{allton}. In addition we perform a
reweighting in $\beta$ by the usual multi histogram reweighting, which is
mandatory to follow the critical line and ensures the overlap of the
configurations. Multi-parameter reweighting was first applied to the problem of
finite density QCD in \cite{fodor}.

\section{THE $m$ DEPENDENCE}
\begin{figure}[htb]
\includegraphics[width=7.5cm]{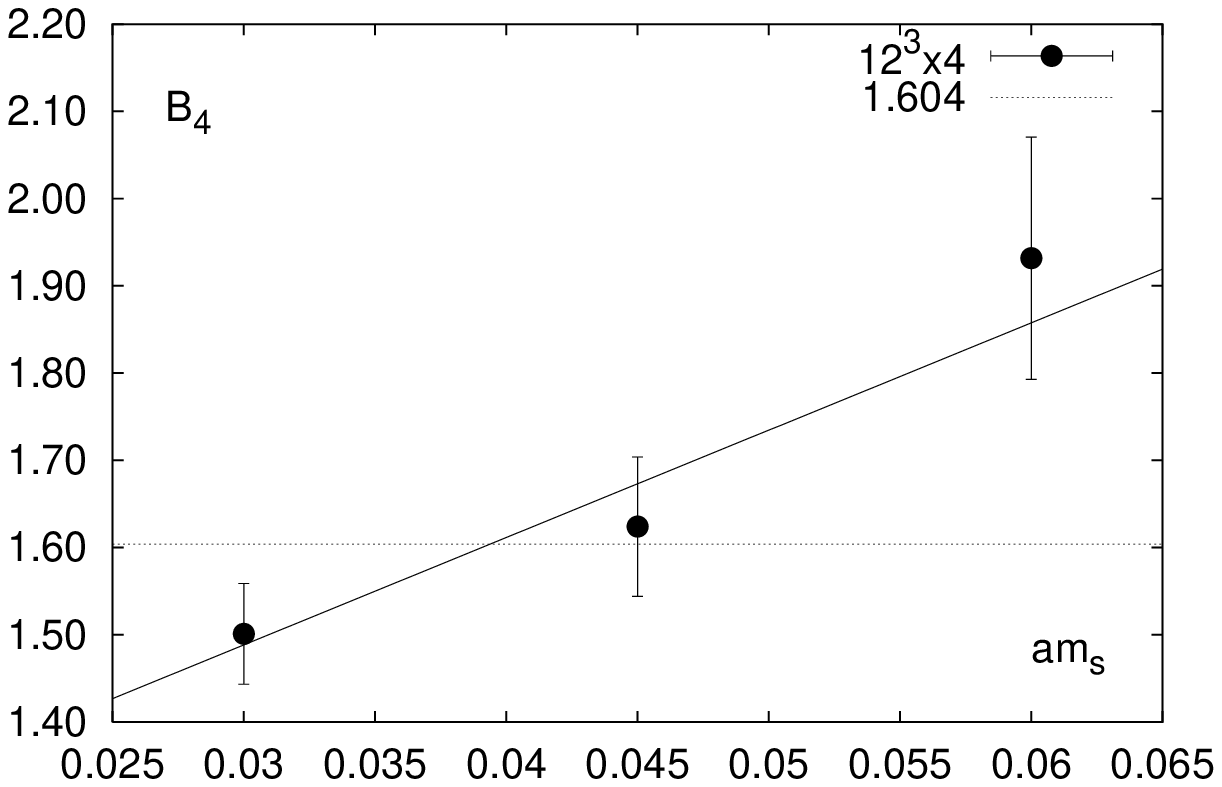}
\includegraphics[width=7.5cm]{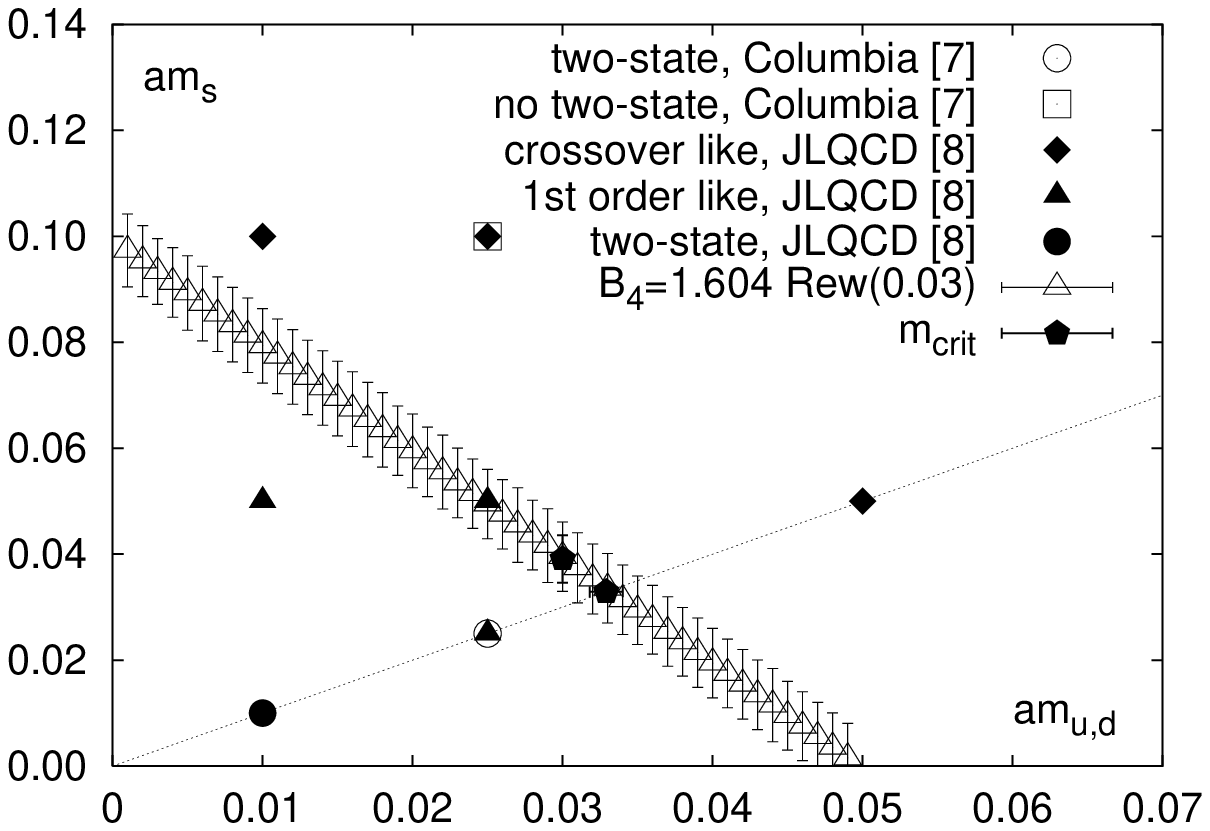}
\caption{(a) Binder Cumulant of (2+1)-flavor QCD for the standard action and
  $am_{u,d}=0.03$. The 3-flavor point is from \cite{schmidt}. (b) Line of
  second order phase transitions in the quark mass plane. }
\label{fig:B4_2+1}
\end{figure}

The 3-flavor critical point for standard staggered fermions was determined in
\cite{schmidt}. To estimate the dependence of this critical point on
non-degenerate quark masses, we performed 2+1 flavor simulations for fixed
$am_{u,d}=0.03$ and $am_s=0.045, 0.06$. The volume is $12^3\times 4$. For each
mass point we have three $\beta$-values. The total number of trajectories is
$52000$ for $am_s=0.045$ and 7300 for $am_s=0.06$. The results for the Binder
Cumulants are shown in figure~\ref{fig:B4_2+1}a.
The straight line fit of $B_4$ together with the assumption of the 3d-Ising
universality class, yields a critical strange quark mass. The critical values of
the 2+1, and 3-flavor\cite{schmidt} calculations are listed in
table~\ref{table:1}.
\begin{table}[b]
\newcommand{\cc}[1]{\multicolumn{1}{c}{#1}}
\newcommand{\rr}[1]{\multicolumn{1}{c|}{#1}}
\begin{center}
\begin{tabular}{|c|lll|} \hline
flavor & \cc{$m_{u,d}$} & \cc{$m_s$}   & \rr{$\beta_{crit}$} \\ \hline
  3    & $0.0329(11)$   & $0.0329(11)$ & $5.1454(5)$         \\
2+1    & $0.0300$       & $0.0391(50)$ & $5.1466(44)$        \\ \hline
\end{tabular}
\end{center}
\caption{Critical points in the quark mass plane (standard action). The
  3-flavor result is from \cite{schmidt}}
\label{table:1}
\end{table}

For mass reweighting the operator $R_1$ is the chiral condensate. We use the
3-flavor data of mass value $am=0.03$ to compute the line of constant
$B_4=1.604$ in the quark mass plane. The result is shown in
figure~\ref{fig:B4_2+1}b. The result is a straight line due to the reweighting
order, for higher order reweighting one would expect deviations from a straight
line. The validity range of the straight line is not known; nevertheless it is
in agreement with the calculations of \cite{columbia,JLQCD}. The slope is $-2$
within errors, which is the expected value in the vicinity of the 3-flavor
critical point.

\section{THE $\mu$ DEPENDENCE}
\begin{figure}[tb]
\includegraphics[width=7.5cm]{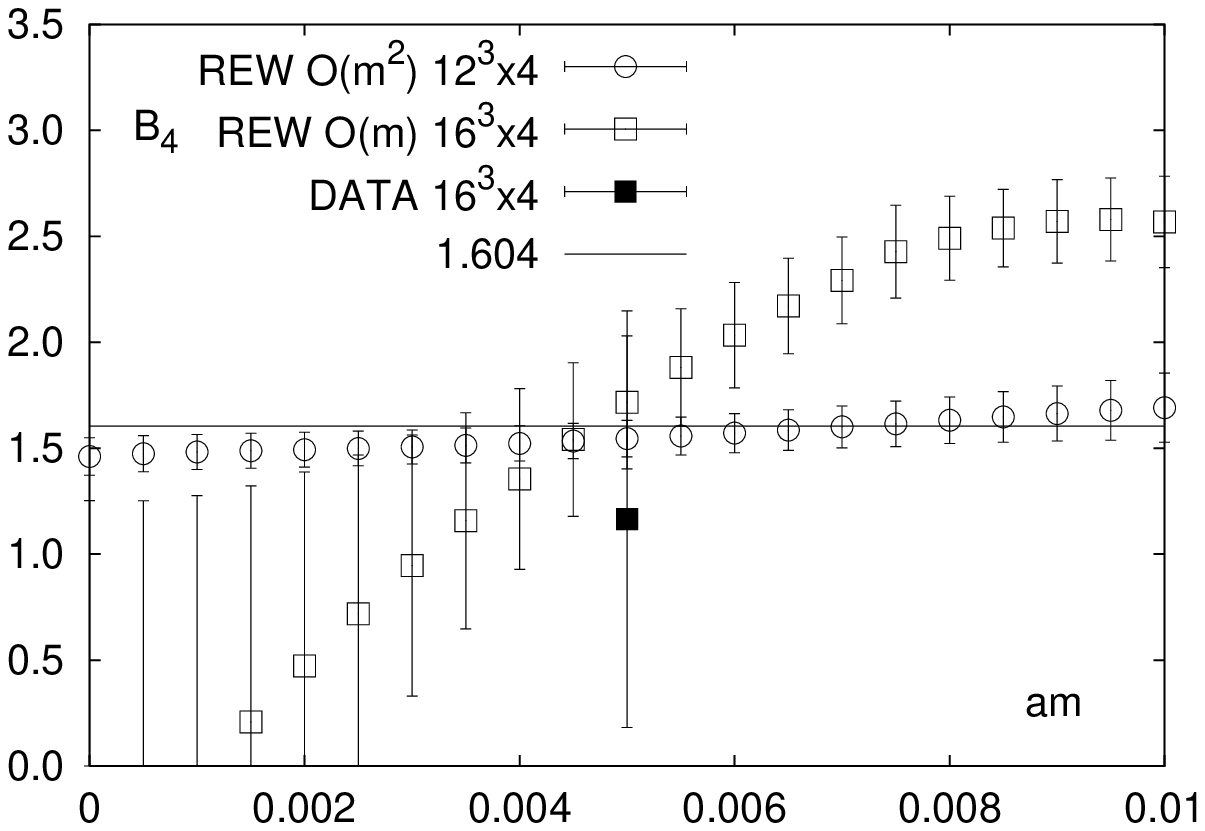}
\includegraphics[width=7.5cm]{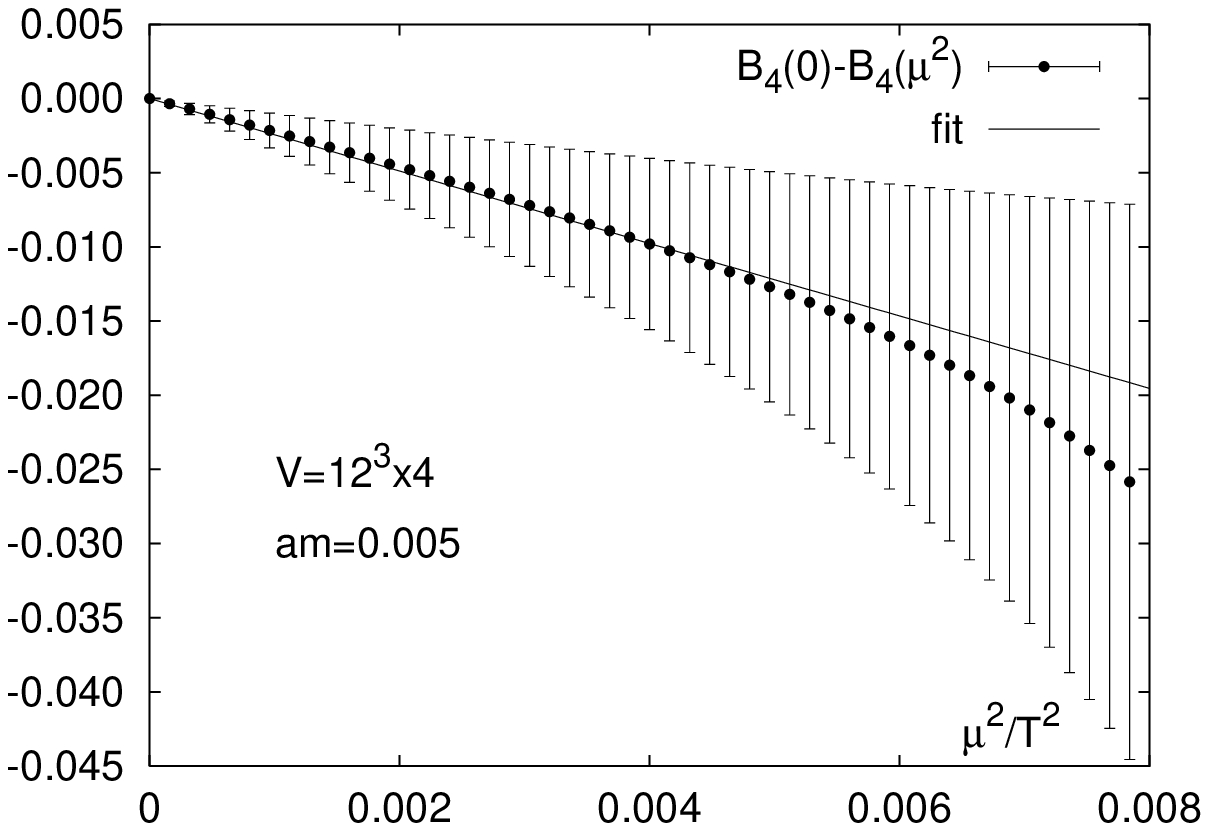}
\caption{Binder Cumulant of 3-flavor QCD for the p4-improved action, (a) as
  a function of $m$, (b) as a function of $\mu^2/T^2$.}
\label{fig:B4_3fp4}
\end{figure}

For 3-flavors of p4-improved fermions, a mass value of $am=0.005$ and
a volume of $12^3\times 4$, we measured the reweighting operators $R_1, R_2$
for $m$ and $\mu$ reweighting. Here we have five $\beta$-values, with a total
number of 6100 trajectories. For $am=0.01$ and $V=16^3\times 4$ we use
$R_1=\bar\psi \psi$ for mass reweighting. To determine the critical mass value
$\bar m$, we compute $B_4$ as a function of $m$. The result is shown in
figure~\ref{fig:B4_3fp4}a.
The two different volumes have an intersection point near the value of
$B_4=1.604$, i.e. 3d Ising universality class. Due to the large
errors we give an upper bound for the critical mass only, which is $a\bar
m<0.0075$, or in terms of the pion mass $\bar m_{PS} < 190 MeV$.

\begin{figure}[tb]
\includegraphics[width=7.5cm]{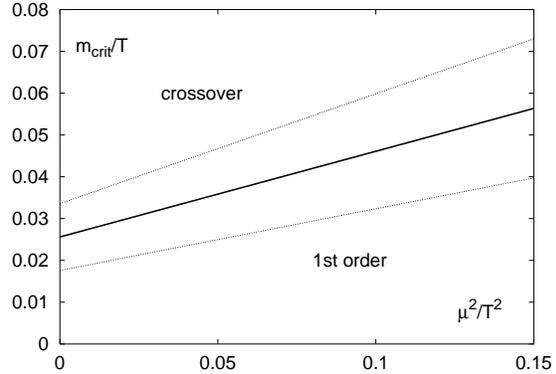}
\caption{Sketch of the phase diagram. Dotted lines indicate the error band.}
\label{fig:sketch}
\end{figure}

For increasing quark chemical potential $\mu_{u} = \mu_{d} = \mu_s$ we find a
decreasing Binder Cumulant, as shown in figure~\ref{fig:B4_3fp4}b for the
$12^3 \times 4$ lattice. From the two partial derivatives $\partial B_4/\partial
(am)$ and $\partial B_4/\partial (a^2\mu^2)$, which we get from straight line
fits of the reweighted data, and the assumption that $\partial B_4/\partial
(am)$ is constant in $am$, one can compute the quantity $\partial (a\bar
m)/\partial (a^2\mu^2)$. The first derivatives $\partial B_4/\partial (a\mu)$
and $\partial (a\bar m)/\partial (a\mu)$ vanish because of symmetry reasons. A
jackknife analysis yields $a^{-1}\partial \bar m/\partial (\mu^2)=0.82(23)$, or
equivalent $T\partial \bar m/\partial (\mu^2)=0.21(6)$. The resulting phase
diagram is sketched in figure~\ref{fig:sketch}. The straight line corresponds
to the curved dashed line in figure~\ref{fig:phase_diagram}.

\end{document}